\documentstyle[12pt]{article}
\begin{document}
\baselineskip .3in
\begin{titlepage}
\begin{center}{\large{\bf On Some Properties of Exotic Baryons in Quasi particle diquark Model}}
\vskip .2in A.Bhattacharya$^{\dag}$, B.Chakrabarti, S.Mani and
S.N. Banerjee.
\end{center}
\vskip .1in
\begin{center}
Department of Physics, Jadavpur University \\
Calcutta 700032, India.\\
\end{center}
\vskip .3in

\vskip .3in Abstract: The properties of the $\Theta^{+}$ and
$\Xi^{--}_{3/2}$ exotic baryons have been investigated in the
framework of diquark- diquark-antiquark system. A model for
diquark suggested in an analogy with the quasi-particle in a
crystal lattice has been used. The mass of the exotics have been
computed. The Decay width of the $\Theta^{+}$ has been estimated
to be $\simeq$ 11 MeV using effective diquark theory. The radius
of the $\Xi^{--}_{3/2}$ has been extracted as 0.94fm using the
experimental value of the
$\frac{\Gamma_{\Theta^{+}}}{\Gamma_{\Xi^{--}_{3/2}}}$. The results
obtained are found to be in good agreement with the present
experimental and theoretical suggestions.

Pacs: 12.39 Ki; 13.30Eg;12.38Lg


\vskip 1in

$^{\dag}$e-mail: pampa@phys.jdvu.ac.in
\end{titlepage}
\newpage
 The theoretcal prediction of pentaquark has been introduced in the framework of the chiral
 soliton model. DPP [1] used the topological solution in
 SU(3)*SU$(\overline{3})$ chiral symmetric sigma model to study the pentaquark state. With
 $N_{c}$ = 3, the SU(3) skryme model led to the prediction of
 SU(3)$\overline{10_{f}}$ which are not composed of three quarks
 but exotics with spin parity restricted to $\frac{1}{2}^{+}$. DPP
 have estimated the mass of the $\Theta^{+}$ baryon as 1540 MeV
 which is experimentally supported by LEPS Group [2] and others
 [3]. However the present experimental status of the $\Theta^{+}$
 is somewhat confusing but it can be asserted that if $\Theta^{+}$
 baryon exists it should have very narrow state. The experimental
 upper limit  is $\Gamma\langle$ 25 MeV whereas
 some experiments [4] suggests the value to be $\Gamma\langle$ 9
 MeV but the phase analysis of KN scattering results in the even
 stronger limit on $\Gamma\langle 1$MeV [5]. Edimuller et al have
 [6] have investigated $\Theta^{+}$ decay width in the QCD sum
 rule and suggested that the color exchange between meson like and
 baryon like cluster reduced the coupling constant four times for
 negative baryons. It has been argued that the pentaquark decay width is
 strongly suppressed due to the fact that the size of the pentaquak
 is no larger than the usual hadronic size [7]. Hong et al [8] have suggested
 that the small decay width is largely due to the large tunneling
 suppression of a quark and a pair of diquarks.In the present paper
 we have estimated the masses and decay width of the exotics using
 the quasi particle model of diquark suggested by us [9].

 It is now growing idea that the deeply bound diquarks are the building
blocks in the formation of the multi quark states. Some
experimental facts like hadron jet formation etc demands the
prediction of diquark. To estimate the mass of the pentaquark
state Jaffe and Wilczek [10] used the diquark model whereas in the
KL model [11] a combination of diquark and triquark is considered.
A number works have been done towards the understanding of the
structure of diquark. The possibility of forming quark-quark and
quark-antiquark system by Instanton Induced Interaction(III) have
been developed by Shuryak [12] and Schafer et.al [13]. Recently we
have suggested a model of diquark in the framework quasi particle
system in  an analogy with the quasi particle similar to an
electron in the crystal lattice [9,14]. Electrons behave as quasi
particle in crystal lattice and such single particle excitations
are termed as quasi particle and the effect of the crystal field
is contained in those quasi particles. The electrons in crystal is
in a situation exactly the same as the elementary particle in
vacuum [15]. Under the influence of combined force of crystal
field ($\nabla V$) and an external force (F) which accelerate the
electron the mass of the electron in a crystal is modified and
behaves like a quasi particle whose effective mass $m^{*}$
reflects the inertia of electrons which are already in a crystal
field such that:
\begin {equation}
m^{*}\frac{dV}{dt}= F_{h}
\end{equation}

and the bare electrons ( with normal mass) are affected by the
lattice force -$\nabla V$,  which corresponds the periodic crystal
potential V as well as the external force F. so that:
\begin{equation}
m\frac{dV}{dt}= F - \frac{dV}{dx}
\end{equation}
 Hence the ratio of the normal mass (m) to the effective mass ($m^*$) can
expressed as:
 \begin{equation}
 m/m^{*}= 1 - \frac{1}{F}[\frac{\delta \overline{V}}{\delta x}]
 \end{equation}
We assumed a similar type of picture for diquark. To get diquark
effective mass inside the pentaquark baryons we assumed that the
 diquark is an independent body which is under the influence of one
 gluon exchange type of field due to the meson cloud represented by potential
 $\overline{V}$
 =-(4/3)$\alpha_{s}$/r in analogy with the crystal field on a crystal
 electron.
 and an average of lattice force F = -ar as  an oscillatory
 external force so that the the ratio of the constituent mass and the
 the effective mass of the diquark inside a hadron may be
 represented as:
 \begin{equation}
 \frac{m}{m^{*}}= 1+ \frac{\alpha_{s}}{kr^{3}}
 \end{equation}
 Here m represents the normal constituent mass of the diquark and
 m$^{*}$is the effective mass of the diquark,$\overline{V}$ being the average value of the
 one gluon exchange potential. With $\alpha_{s}$ = .58 [16],
 k=241.5MeV [17], $m_{u}$ = $m_{d}$ = 360MeV, radius parameter of ud diquark as = 5.38 $GeV^{-1}$ [18]
 which approximately represents the pion radius, the effective mass $M_{ud}$ of
 the ud diquark has been estimated to 506 Mev. The diquarks which are described as the
 elementary excitation simulating many body interaction behaves
 like scalar bosons. Assuming that such a system of low lying excitations
 behaving like quasi particles, do not interact among themselves
 and their energies are additive [19], the mass of the
 $\Theta^{+}$ baryon has been estimated as:
 \begin{equation}
 \Theta^{+}= 2M_{ud}+ m_{s}
 \end{equation}
 which yields $M_{\Theta^{+}}$ = 1552MeV.
 Similarly considering the $\Xi^{--}_{3/2}$
 as(sd)(sd)$\overline{u}$ system the mass can be estimated from
 the mass of the (sd) diquark using the equation (4). With the
 constituent mass $m_{d}$+ $ m_{s}$ = 900 MeV and the sd diquark radius
 parameter $\simeq$ 6.91 $GeV^{-1}$ [20] which is
 approximately equals the kaon radius, the [sd] diquark mass is obtained as 751.25 MeV  so that  the
 mass of the
 $\Xi^{--}_{3/2}$ is estimated as:
\begin{equation}
 \Xi^{--}_{3/2}= 2M_{sd}+ m_{u}
 \end{equation}
 which yields $M_{\Xi^{--}_{3/2}}$ = 1862.25 MeV. From (5) and (6) we
 get:
\begin{equation}
M_{\Xi^{--}_{3/2}}-M_{\Theta^{+}}=(2M_{sd}+ m_{u})-(2M_{ud}-m_{s})
=310.25MeV.
 \end{equation}
 which agrees reasonably well to the splitting estimated from
 NA49 [21] experiment. The masses of the exotic baryons estimated
 are in well within experimental limits.

  {\bf Decay Properties of $\Xi^{--}_{3/2}$, $\Theta^{+}$}:

   To estimate the decay width we consider the process followed by
   Hong et al [8]. In the context of the chiral effective theory
   it is assumed that a 'd' quark tunnels from a diquark to form a
   ud to other diquark to form a udd and a off-shell u to be annihilated by anti strange quark.
   Decay width can be expressed as:
    \begin{equation}
    \Gamma_{\Theta^{+}}(\Theta^{+}\rightarrow K^{+}+n)
    \simeq 5e^{-2s_{0}}[g^{2}g_{A}^{2}/8\pi f_{k}^{2}]|\psi(0)|^{2}
    \end{equation}
    Using the WKB approximation $e^{-s_{0}}$ $\simeq$
    $e^{-\Delta E.r_{0}}$ where $\Delta E$ = $m_{u}$+ $m_{d}$ - $M_{D}$ is diquark binding
     energy, $r_{0}$ is [ud] diquark radius which is taken as 5.38 $GeV^{-1}$.  $|\psi(0)|^{2}$ is quark-diquark wave function. To
     get an estimate of the $\Gamma_{\theta^{+}}$ we need the wave
     function $|\psi(0)|^{2}$ for the quark-diquark system. We
     assumed that the pentaquark baryon is described by the hadron
     wave function obtained from the statistical model [22] which
     runs as:
     \begin{equation}
     |\psi(0)|^{2} =315/64 \pi r_{\Theta^{+}}^{3}
     \end{equation}
     where $r_{\Theta^{+}}^{3}$ is the radius parameter of the quark-diquark
     system and may be approximated by the pentaquark
     radius. Recently Panda et al [23] have investigated the
     properties of the $\Xi^{--}_{3/2}$, $\Theta^{+}$ baryons in
     the symmetric quark matter in the quark-meson coupling model.
     In course of investigating the effect of mass on exotics they
     have estimated the ratio of the radius of $\Theta^{+}$ and nucleon
      $\frac{R_{\Theta}}{R_{N}}$. They have estimated
      $r_{0}$ ($\Theta^{+}$) $\sim$3.41fm fixing nucleon radius at
      .6fm. Ellis et al [24] have estimated $\Theta^{+}$ radius as $\sim$ 1/248 MeV = .80fm.in
      the context of two dimensional QCD whereas Kahana et al [25] have considered the radius
      parameter of $\Theta^{+}$ as $\sim$1.13fm = 5.65 $GeV^{-1}$ in the context of
      describing the $\Theta^{+}$ as kaon-nucleon resonance at 100MeV having
      width 7-21MeV with a phenomenological effective potential.
      With the input of $r_{0}\Theta^{+}$ from [25] we have
      estimated the $|\psi(0)|^{2}$ for
      $\Theta^{+}$=.0086$GeV^{3}$. The binding energy for the
      ud diquark is obtained as $\Delta$ E = $m_{u}$+ $m_{d}$- $M_{ud}$ =
      720-506=214MeV in the present work. With the input  of $\Delta$ E, $|\psi(0)|^{2}$
      estimated above and $f_{k}$=160 Mev [26], $g_{A}$ =0.75, $g^{2}$ = 3.03
      in the expression (8) we have obtained the decay width
       $\Gamma_{\Theta}^{+}$ $\simeq$11.01MeV.

       The decay width of the $\Xi^{--}_{3/2}$ baryon may be
       expressed as [8]:
       \begin{equation}
    \Gamma_{\Xi^{--}_{3/2}}
    \simeq .5e^{-2s_{0}}[g^{2}g_{A}^{2}/8\pi f_{\pi}^{2}]|\psi(0)_{\Xi^{--}_{3/2}}|^{2}
    \end{equation}
    where $S_{0}$ = $\Delta$E.$r_{0}$. $\Delta$E
    is the binding energy of ds diquark which is estimated in the
    present work as $\Delta$E = $m_{d}$+ $m_{s}$- $M_{ds}$ =
    900-751.25 = 148.75MeV. $r_{0}$ is the radius of the [ds] diquark which is assumed to be
     6.75$GeV^{-1}$ as stated before. Let $r_{\Xi}$ be the radius
    parameter for $\Xi$, the wave function for the $\Xi^{--}_{3/2}$ may be
    expressed as before in the context of the statistical model
    as:
    \begin{equation}
     |\psi(0)|^{2} =315/64 \pi r_{\Xi}^{3}
     \end{equation}
     so that the from (10) with $f_{\pi}$ = 92MeV we arrive at:
     \begin{equation}
    \Gamma_{\Xi^{--}_{3/2}}=.848/r_{\Xi}^{3}
     \end{equation}
     From the expression (8) and (12)we get:
     \begin{equation}
      \frac{\Gamma_{\Xi^{--}_{3/2}}}{\Gamma_{\Theta^{+}}}= 77.09/r_{\Xi}^{3}
      \end{equation}
      The relative partial decay width of $\Xi^{--}_{3/2}$, $\Theta^{+}$ have
      been reported by CERN SPS and LEPS as [8] $\sim $.7. With
      this experimental input of
      $\frac{\Gamma_{\Xi^{--}_{3/2}}}{\Gamma_{\Theta^{+}}}$, the
      radius of the $\Xi^{--}_{3/2}$ is estimated as
      4.72$GeV^{-1}$= .94fm. Hong et al [8] have studied the properties of exotics in the
      chiral effective theory and obtained $\Gamma_{\Theta^{+}}$ and
      $\Gamma_{\Xi^{--}_{3/2}}$ as 2.5 $\sim$ 7MeV
      and 1.7 $\sim$ 4.8 respectively considering flavour
      independent binding energy of diquark as 270MeV.

      {\bf Conclusions}

  In the present work we have estimated the mass and decay width of the exotic
             baryons considering it as diquark-diquark-antiquark
             system. In the quasi particle model a diquark is considered to be an elementary
             excitation simulating the many body interaction such
             that they behave like scalar boson, a highly
             correlated spin zero particle [qq]I=0.  We have
             studied the properties of $\Xi^{--}_{3/2}$,
             $\Theta^{+}$in the context of the proposed quasi
             particle model of diquark and obtained good agreement
             with the existing theoretical and experimental
             findings [8]. It may be noted that the wave function for hadrons used from the
             statistical model reproduces reasonable results for exotics
             also. However the radius parameters of the diquark and pentaquarks
              used in the present investigation
             are not exactly known but are found to be well approximated by corresponding
             meson and
        baryon radius parameter estimated in the context of the statistical model [18,20].
              Recently Maris et al [27] have
             calculated [ud] and [us] diquark charge radius as
             0.51$fm^{2}$ and .63/or.71$fm^{2}$ respectively and
             suggested that $r_{\Theta^{+}}$  $\simeq$1fm. We have
             estimated  radius of the $\Xi^{--}_{3/2}$ as .94
             fm. The masses of the exotics (pentaquarks) estimated in the present work are found to be
             very close to the experimental value. Most important work is now to understand
             the true nature of the diquark which are suggested to be  the fundamental building
             blocks of nature [28]. The quasi particle model for diquark appears to
             be very successful in reproducing some important
             results of exotics and hadrons.

\newpage

\noindent

[1] D.Diakonov et al., Z.Phys. A {\bf 359}(1997)305.

\noindent
 [2]  T.Nakano et al., Phys. Rev. Lett. {\bf 91}(2000)
262001.

\noindent
 [3] S.Stepanyan et al.,(CLAS Collaboration),

Phys.Rev.Lett.{\bf 91}(2003)252001;J.Berth et.al.,(SAPHIR

Collaboration), Phys.Lett.B {\bf 572}(2003)127

\noindent
[4] V.V.Berrnin et al., Yad.Fiz.(2003)1763;
Phys.At.Nucl.{\bf 66}(2003)1715.

\noindent
[5] R.L. Jaffe et al., Phys.Rev.D{\bf 71}(2005)034012;
R.N.Cahn et.al., Phys.rev D{\bf 69}(2004) 011501.

\noindent [6] M.Eidemuller et al., Phys. Rev.D{\bf
72}(2005)034003.

\noindent
 [7] B.L.Ioffe et al., JEPT Lett {\bf 80}(2004)386.

\noindent [8] D.K.Hong et al.; hep-ph/0403205.

\noindent
 [9] B. Chakrabarti et al.,5th International Conference
on Perspective in Hadronic Physics at ICTP, May(2005)(to be
published in Nucl.Phys.A as procedings Suppl.)

\noindent [10] R.L.Jaffe et al., Phys. Rev. Lett.{\bf 91}(2003)55.

\noindent [11]M.Karlinar et al., Phys. Lett.{\bf 575}(2003)249.

\noindent
[12] E.V.Shuryak., Nucl. Phys. {\bf B50} 93(1982)

\noindent
[13]T.Schafer et al., Rev. Mod. Phys.{\bf 70}(1998) 323.

\noindent
[14] A.Bhattacharya. et al., hep-ph/0604221.

\noindent [15]  A.Haug; Theoretical Solid State
               Physics,Pergamon Press; 100(1972)

\noindent [16]  W.Lucha et al., Phys.Rep.{\bf 200} 127(1991)

\noindent [17] R.K.Bhaduri et al., Phys. Rev. Lett.{\bf 44}
               (1980)1369

\noindent [18] S.N.Banerjee et al.; Int.J Mod.Phys.A{\bf
               2} (1987)1829.

\noindent [19] L.D.Landau and E.M.Lifshitz.; 'Statistical
                 Physics'Vol.5,3rd edition.Part I (Butterworth Heinemann,
                 Moscow,1997)p216

\noindent [20]S.N.Banerjee et al.,Had.J.{\bf 12}(1989); S.Wan et
al.,Z.Phys.C{\bf 73}(1996)141.

\noindent [21] A.Alt et al.,(NA49 Collaboration)'
Phys.Rev.Lett.{\bf 92}(2004)042003.

\noindent [22] S.N.Banerjee et al.,Int.J.Mod.Phys.A{\bf
16}(2001)201; Int.J.Mod.Phys.A{\bf 17}(2002)4939.

\noindent [23] P.K.Panda et al.,Phys.Rev.c{\bf 72}(2005)058201.

\noindent [24] J.Ellis et al.,JHEP{\bf 08}SISSA (2005)

\noindent [25] D.E.Kahana et al.,Phys.Rev.D{\bf 69}(2004)117502.

\noindent [26] J.F.Amemdson et al., Phys. Rev.D{\bf 47}(1993)3059.

\noindent [27] P.Maris , Few Body. Syst 32 (2002) 41;
Nucl-th/0204020

\noindent [28]M.Oka, Prog.Ther.Phys.{\bf 112}(2004)1;R.L.Jaffe.
Phys.Rep.{\bf 409}(2005)1.

\end{document}